\documentclass[twocolumn,preprintnumbers,amsmath,amssymb]{revtex4}
\usepackage{epsfig}
\usepackage{graphicx}% Include figure files
\usepackage{dcolumn}% Align table columns on decimal point
\usepackage{bm}% bold math

\begin{document}

%\preprint{APS/123-QED}

\title{Nanolithography using counter-propagating light-sheets}% Force line breaks with \\

\author{Kavya Mohan and Partha Pratim Mondal }
 %\altaffiliation{}%Lines break automatically or can be forced with \\
 %\email{partha@fisica.unige.it}

\affiliation{%
Nanobioimaging Laboratory, Department of Instrumentation and Applied Physics, Indian Institute of Science, Bangalore 560012, INDIA
}%

\date{\today}% It is always \today, today,

\begin{abstract}
We propose a laser interference nano-lithography technique for fabrication of nano-structures. This is inspired by a $2\pi$-illumination system that consists of two cylindrical lens arranged face-to-face at a distance $2f$ with a common geometrical focus. When illuminated by a coherent light, this results in the super-position of two counter-propagating light-sheets. The interference gives rise to nano-bump structures along the optical axis. This technique overcomes the existing point-scanning techniques and paves the way for mass production of nano-structures. Study shows the structures with a feature size of $60~nm$ and an inter-nanobump separation of $180~nm$. Proposed technique may find applications in plasmonics, nanophotonics and nanobiology. 
\end{abstract}

\maketitle

%PACS: 87.64.mn, 87.64.kv, 87.85.Pq \\

Light-sheet microscopy has seen explosive progress in the last decade and it is finding applications in fields ranging from applied physics to nano-biology \cite{voie1993} \cite{huisken2004}\cite{kavya2014}. This techniques was first realized by Siedentopf et al. \cite{siedentopf1903} and the first imaging application was demonstrated by Voie et al. \cite{voie1993}. Thereafter, the technique was vastly developed by Stelzer group and exploited for potential applications in Biology \cite{huisken2004}\cite{keller2010}. Many variants of light-sheet microscopy have emerged over the last few years. This include, thin light-sheet microscopy \cite{fuchs2002}, ultramicroscopy \cite{dodt2007}, digital light-scanning microscopy (DLSM) \cite{keller2008}, objective coupled plane illumination microscopy (OCPI) \cite{holekamp2008} and confocal light-sheet microscopy \cite{silv2012}\cite{baum2012}. Over the years, light-sheet based techniques have aided many biological studies \cite{dodt2007}\cite{ahrens2013}\cite{truong2011} \cite{fra2011}\cite{raju2013}. There are many advantage of light-sheet microscopy over point-scanning techniques  and holds a lot of benefit as far as imaging and nano-lithography are concerned. What make light-sheet illumination so special is the large field-of-view and the ability to scan the entire plane in a single-shot. This is very different from the existing point-illumination techniques that require sequential scanning. Light-sheet illumination has the ability of patterning selective plane over a large area with better accuracy and high reproducibility. \\ 

Nano-lithography techniques can be broadly divided into two major types i.e, masked and maskless. Fabricated masks and moulds are used to transfer pattern over a desired template for masked lithography technique, while maskless lithography employ some kind of beam (ion beam, electron beam or light beam). Masked lithography has the advantage of high-throughput fabrication over a large area, while maskless techniques use spot size of the beam for precise patterning. Popular masked nano-lithography technique include, nano-imprint lithography \cite{chou1996}\cite{cattoni2010}, photolithography \cite{wagner2010}\cite{sanders2010} and soft-lithography \cite{xia1998}\cite{cooper2011}. On the other hand, maskless lithography include, focused ion-beam lithography \cite{mel1987}\cite{tseng2004}, electron beam lithography \cite{alt2010}\cite{gri2009}, scanning probe lithography \cite{piner1999}\cite{shim2011} and two-photon direct laser writing \cite{wollhofen2013}\cite{burmeister2012}\cite{dong2008} \cite{denk1990}. It is the slow serial nature of maskless technique that makes it inappropriate for mass production. So, there is a trade-off between high-throughput production and high precision reproducibility. Ultimately, existing techniques are limited by Abbe's diffraction limit and thus cannot generate feature smaller than this limit \cite{abbe1884}\cite{hell2007}.  Recently, it has been shown  that, diffraction-unlimited nano-lithography can be achieved by using STED technique \cite{ben2012}\cite{wollhofen2013}. Of-late, high-throughput nano-fabrication using visible light is in demand due to the advancement in nano-technology for enabling mass production of nano-devices.  \\ 

In this letter, we propose a light-sheet based illumination technique for nano-lithography. The optical configuration consists of two cylindrical lens placed opposite to each other and bear a common geometrical focus. We termed this technique as $2\pi$-system since the aperture angle of combined cylindrical lens system can occupy a maximum angle of $2\pi$. The motivation behind the proposed geometry is to generate multiple nano-bump pattern. Studies show that the nano-bump pattern has an inter-bump separation of approximately $\lambda /2$. Comparatively, recent developments in laser inference lithography demonstrate an inter-period spacing of $\lambda/2\sin\theta$, $\theta$ being the inclination angle \cite{seo2013}\cite{wang2013}. Proposed technique is a simple and efficient way of generating nano-pattens with high reproducibility. Since, this technique produce periodic pattern, so it has potential applications in photonic crystals and submicron sieves. \\

\begin{figure*}
\includegraphics[height=2.5in,angle=0]{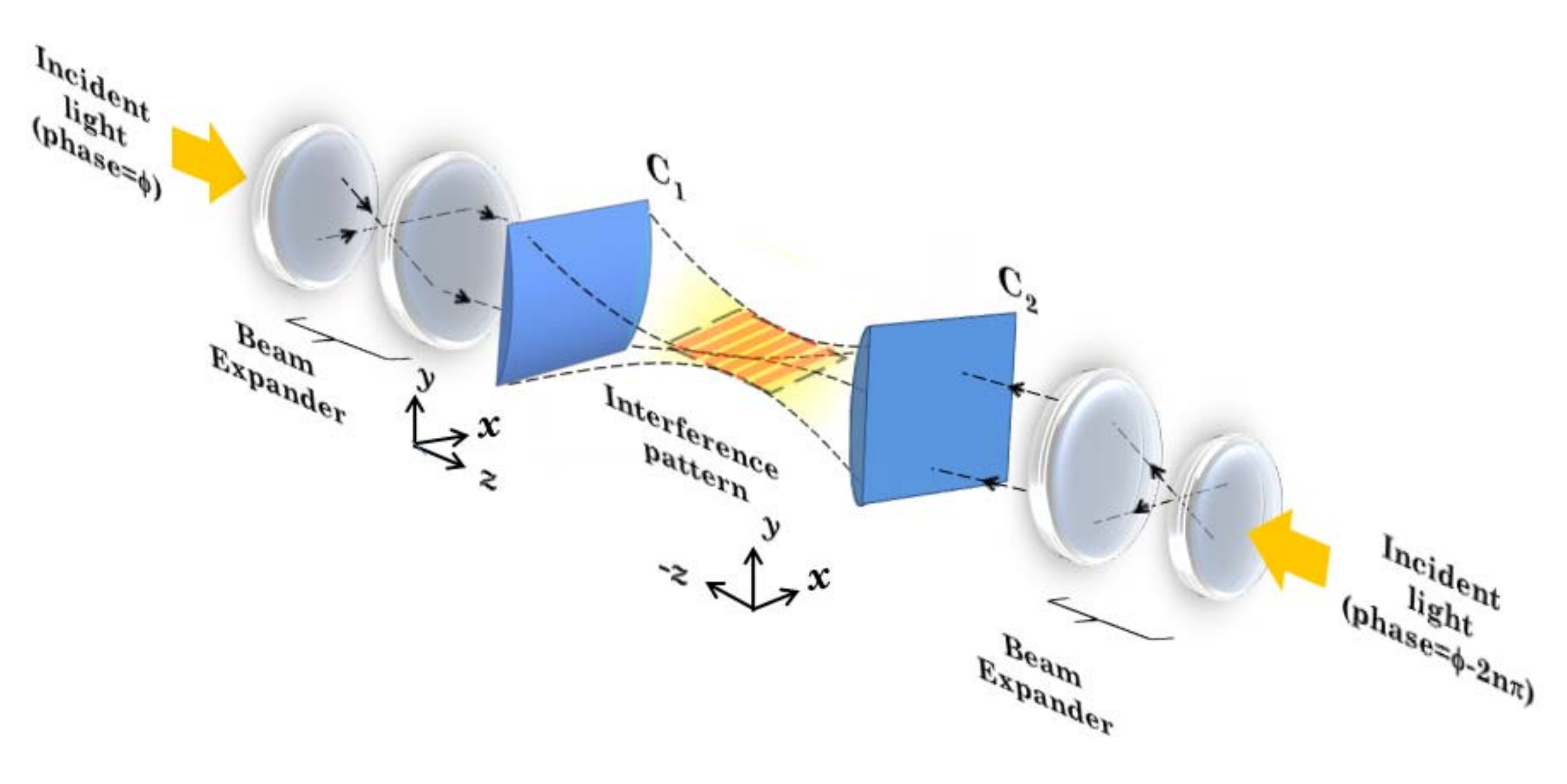}
\caption{\small\bf{Illustration of the proposed light sheet based 2$\pi$ setup which employs two opposing cylindrical lens to produce interference of counter propagating light sheets.}}
\end{figure*}
% Figure ends***********************************

%% Figure begins*********************************
\begin{figure}
\includegraphics[height=3.0in,angle=0]{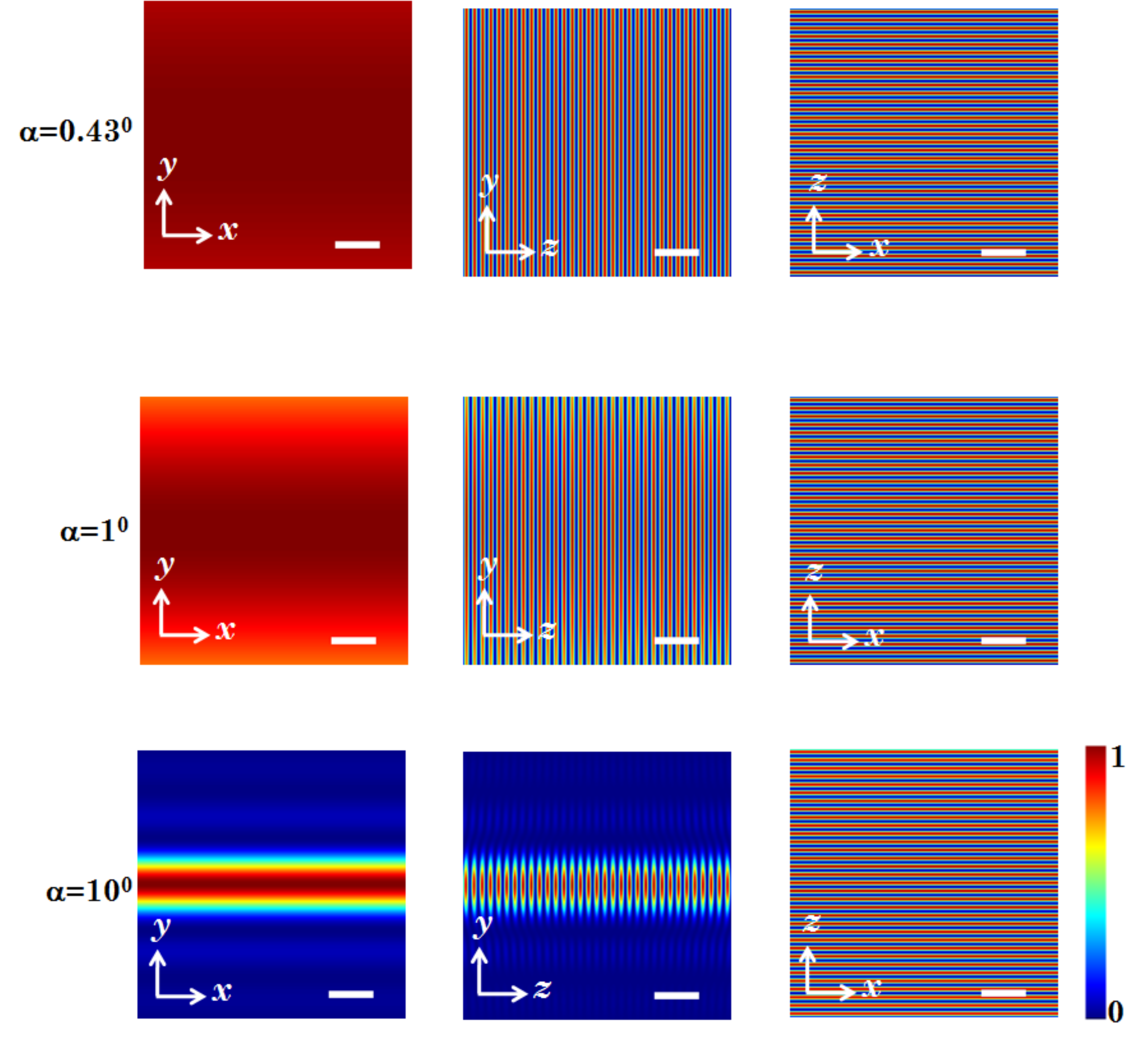}
\caption{\small\bf{xy , yz and xz views of the field distribution calculated for different semi-aperture angle($\alpha$) $\alpha=0.43^o$, $\alpha=1^o$  and  $\alpha=10^o$.Scale bar is 1$\mu m$.} }
\end{figure}
%% Figure ends*********************************** 

The schematic of the optical configuration for creating periodic nano-bump pattern is shown in Fig.1. The setup consists of two opposing cylindrical lenses  $C_1$ and $C_2$ of same focal length (f). The incident coherent beam is allowed to pass through the beam expander to fill the back aperture of the cylindrical lens. When illuminated by the coherent wavefront, both the cylindrical lens give rise to light-sheets which upon superposition generates bump-like interference pattern. The electric field components for a linearly-polarized plane wave at the focus of a cylindrical lens $C_1$ with its axis along x-axis is given by \cite{sub2013},  
\begin{eqnarray*}
\left[\begin{array}{c} e_x(\rho,\phi,z)\\ e_y(\rho,\phi,z)\\ e_z(\rho,\phi,z) \end{array} \right] =  AE_0 \int\limits_{-\alpha}^{+\alpha} ~\left[\begin{array}{c} \cos\theta_p\\ \sin\theta_p\cos\theta\\ \sin\theta_p\sin\theta \end{array} \right] ~\sqrt{\cos\theta}
\end{eqnarray*}
\begin{eqnarray}
\times ~\exp{[i\rho k \cos(\theta-\phi)]} ~d\theta
\end{eqnarray}

\begin{eqnarray*}
\textnormal{where,} A=\sqrt{\frac{fk}{2\pi}} e^{-ifk} e^{i\pi /4}  \sqrt{\frac{n_1}{n_3}} \nonumber
\end{eqnarray*} 
$k$ is the wavenumber in the image space and $\alpha$ is the semi aperture angle. The terms f, $n_1$ and $n_3$ denotes the focal length and refractive index of object and image space respectively. The radial distance from the $x-$axis and the polar inclination are denoted by $\rho=\sqrt{y^2+z^2}$ and $\phi=\tan^{-1}(y/z)$ respectively. \\

For the cylindrical lens $C_2$ the coordinates with respect to lens $C_1$ are $(x,y,-z)$ and the resulting field components are, 
\begin{eqnarray*}
\left[\begin{array}{c} e_x\prime (\rho,\phi,-z)\\ e_y\prime(\rho,\phi,-z)\\ e_z\prime(\rho,\phi,-z) \end{array} \right] =  AE_0 \int\limits_{-\alpha}^{+\alpha} ~\left[\begin{array}{c} \cos\theta_p\\ \sin\theta_p\cos\theta\\ \sin\theta_p\sin\theta \end{array} \right] ~\sqrt{\cos\theta}
\end{eqnarray*}
\begin{eqnarray}
\times ~\exp{[i\rho k \cos(\theta-\phi\prime)]} ~d\theta
\end{eqnarray}

where,$\phi\prime=\pi-\tan^{-1}(y/z)$.\\

The electric field $\vec{E}$ at and near the geometrical focus is the superposition of two counter-propagating electric fields,
\begin{eqnarray}
\vec{E}(\rho, \phi, z) = \vec{E}_1(\rho, \phi, z) + \vec{E}_2 (\rho, \phi, -z) 
\end{eqnarray}

So, the field distribution is given by,
\begin{eqnarray*}
I=|\vec{E}(\rho, \phi, z)|^2 = |\vec{E}_1(\rho, \phi, z) + \vec{E}_2 (\rho, \phi, -z)|^2 
\end{eqnarray*}
\begin{eqnarray}
=|E_x(\rho, \phi, z)|^2 + |E_y(\rho, \phi, z)|^2 +|E_z(\rho, \phi, z)|^2
\end{eqnarray}

where $E_x$,
\begin{eqnarray*}
E_x(\rho,\phi,z) = e_x(\rho,\phi,z)+e_x\prime(\rho,\phi\prime, -z) 
\end{eqnarray*}

\begin{eqnarray*}
 = AE_0 \cos \theta_p \int\limits_{-\alpha}^{\alpha} \sqrt{\cos\theta} \biggl[ e^{i\rho k\cos(\theta-\phi)} + e^{i\rho k\cos(\theta-\pi+\phi)} \biggr] d\theta
\end{eqnarray*}

\begin{eqnarray*}
 = AE_0 \cos \theta_p \int\limits_{-\alpha}^{\alpha} \sqrt{\cos\theta} \biggl[ e^{i\rho k\cos(\theta-\phi)}+e^{-i\rho k\cos(\theta+\phi)} \biggr] d\theta
\end{eqnarray*}

%% Figure begins*********************************
\begin{figure*}
\includegraphics[height=4.5in,angle=0]{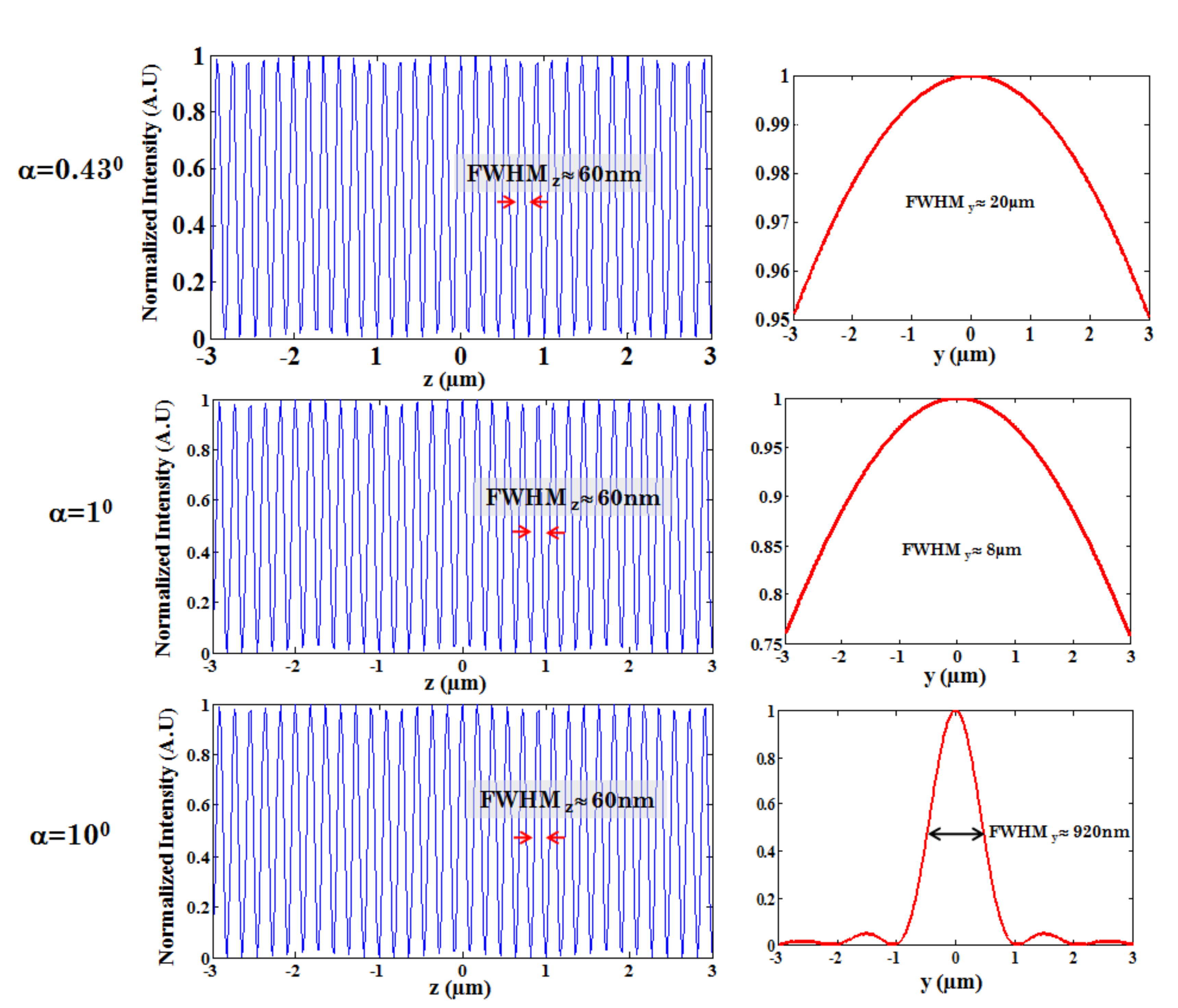}
\caption{\small\bf{Effect of variation of the semi-aperture angle ($\alpha$) on the axial (z-axis) and lateral (y-axis) intensity profile plots. Studies were carried out for $\alpha=0.43^0$, $\alpha=1^0$  and $\alpha=10^0$.It is observed that the changes along y-axis are significant while the pitch along z-axis remains unchanged.} }
\end{figure*}
%% Figure ends*********************************** 

%% Figure begins*********************************
\begin{figure*}
\includegraphics[height=3.3in,angle=0]{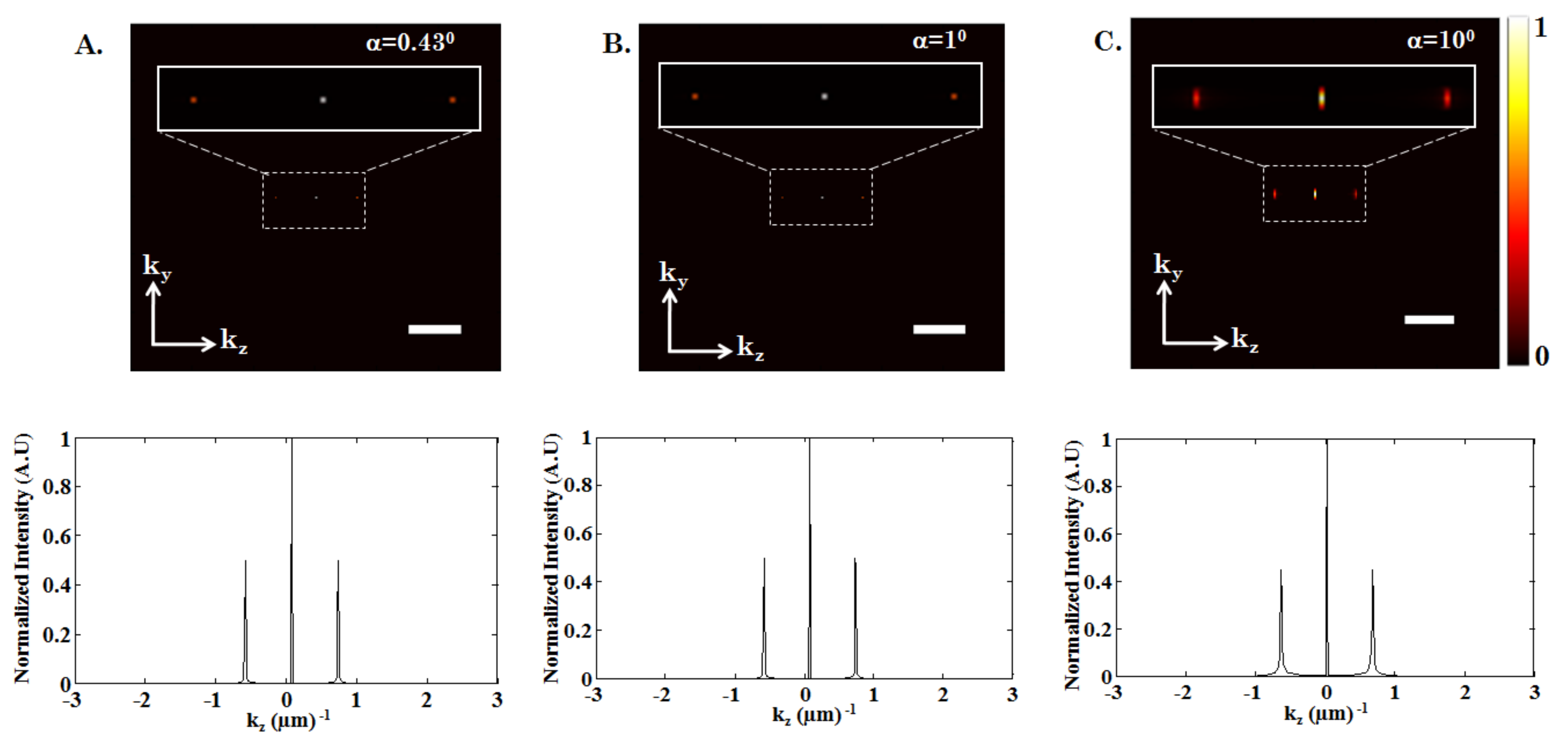}
\caption{\small\bf{A-C shows the OTF of the proposed setup respectively for $\alpha=0.43^0$ , $\alpha=1^0$ and $\alpha=10^0$   and corresponding profile plots  along axial spatial frequency($k_z$)direction. The inset shows the zoomed in view of the OTFs. Scale bar is 1$\mu m$. } }
\end{figure*}
%% Figure ends*********************************** 

Expansion and simplification produces,
\begin{eqnarray*}
 = AE_0 \cos \theta_p \int\limits_{-\alpha}^{\alpha} \sqrt{\cos\theta} ~e^{i\rho k\sin\theta\sin\phi} \biggl[e^{i\rho k\cos\theta\cos\phi}+
\end{eqnarray*}
\begin{eqnarray*}
e^{-i\rho k(\cos\theta\cos\phi)} \biggr] d\theta
\end{eqnarray*}
\begin{eqnarray*}
 = 2AE_0 \cos \theta_p \int\limits_{-\alpha}^{\alpha} \sqrt{\cos\theta} ~e^{[i\rho k\sin\theta\sin\phi]} \cos\gamma  ~d\theta  
\end{eqnarray*}
where, $\gamma = \rho k\cos\theta\cos\phi$. \\

Similarly, the expressions for $y$ and $z$ components of the field can be determined by inspection i.e,
\begin{eqnarray*}
E_y(\rho,\phi,z) = e_y(\rho,\phi,z) +e_y\prime(\rho,\phi\prime, -z) 
\end{eqnarray*}

\begin{eqnarray}
 = 2AE_0 \sin \theta_p \int\limits_{-\alpha}^{\alpha} \cos^{3/2}\theta  ~e^{i\rho k\sin\theta\sin\phi} \cos\gamma ~d\theta  
\end{eqnarray}

and, 

\begin{eqnarray*}
E_z(\rho,\phi,z) = e_z(\rho,\phi,z) + e_z\prime(\rho,\phi\prime, -z) 
\end{eqnarray*}

\begin{eqnarray}
 = 2 A E_0 \sin \theta_p \int\limits_{-\alpha}^{\alpha} \sin\theta \sqrt{\cos\theta} ~e^{i\rho k\sin\theta\sin\phi} \cos\gamma d\theta  
\end{eqnarray}

%% Figure begins*********************************
\begin{figure*}
\includegraphics[height=4.0in,angle=0]{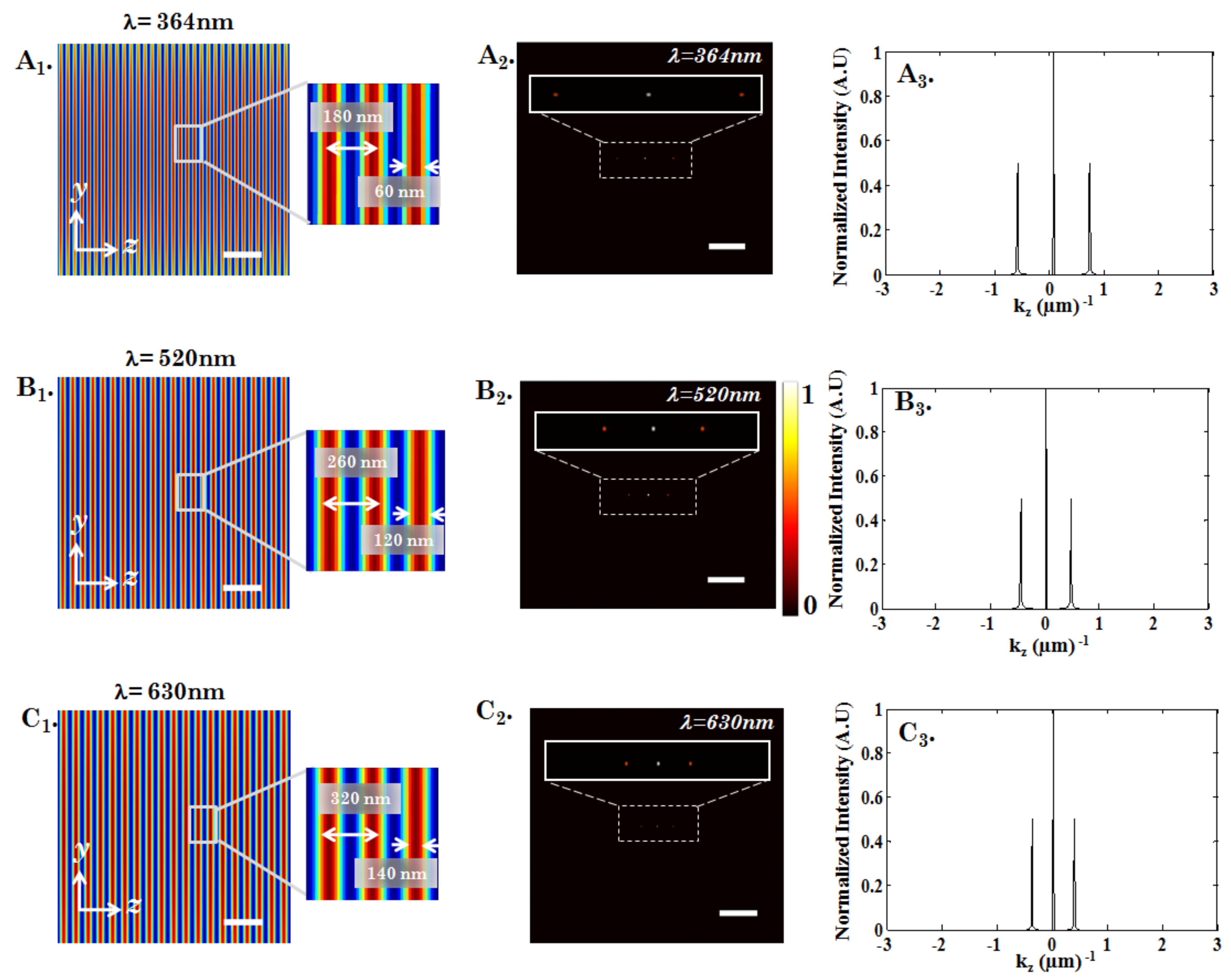}
\caption{\small\bf{ Effect of change in illumination wavelength($\lambda$): $A_1-C_1$  shows the yz view of the field distribution calculated for $\alpha=1^0$ its corresponding OTF is shown in $A_2-C_2$ whereas $A_3-C_3$  shows the intensity profile plot taken through the center of the OTF along the axial spatial frequency ($k_z$ ) direction. Scale bar is 1$\mu m$. } }
\end{figure*}
%% Figure ends*********************************** 

We have carried out computational studies to characterize the field distribution for the proposed light-sheet based $2\pi$-system. Fig.1 shows the schematic diagram of the proposed optical system. The numerical calculations for determining the field distribution at and near the common focus was carried out over a grid size of $300 \times 300 \times 300$ pixels (along, x,y,z axes). Practically, this spans over a spatial dimension of $216~\mu m^3$. An incident light of wavelength $364~nm$ is chosen for the study. The field is computed using eqn.(4) for varying semi-aperture angle ($\alpha$) of cylindrical lens. Fig.2 shows the field distribution along the optical axis (near focus) for different semi-aperture angles. It is apparent that, the thickness of the sheet scales down with increase in aperture angle from $0.43^o$ to $10^o$ (see, Fig. 2, XY-profiles). The nano-bump pattern is evident from the $yz$- and $zx$- planes. Specifically, the field-of-view can be varied along $y$-axis using variable semi-aperture angle. This suggests that such a discrete and dense bump-like pattern can be obtained at nano-scale dimensions. The corresponding intensity plots along axial (z-axis) and lateral (y-axis) axes are shown in Fig.3. From the profile plots it is observed that the z-profile remains unchanged with the change in semi-aperture angle ($\alpha$) while the change along y-axis is significant. Specifically, the $FWHM_y$ along y-axis varies from $1~\mu m$ to $20~\mu m$ for $\alpha=10^o$ to $0.43^o$. We observed that, the separation between two successive maxima is $180~nm$ ($\approx\lambda/2$) whereas, the full width at half maxima ($FWHM_z$) of the individual nano-bump is approximately $60~nm$.\\

To better understand the nano-pattern characteristics, we obtained the optical transfer function (OTF). This provides the information about the spatial frequencies present in the generated nano-bump pattern. The corresponding $k_y k_z$-plane is shown in Fig.4 for semi-aperture angles, $\alpha=0.43^o , ~1^o , 10^o$. One can immediately observe higher spatial frequencies along $k_z$-axis. OTF shows maximum bandwidth along axial direction but it is  not continuous and are separated. Specifically, the low frequency component occur at $k_z=0\mu m^{-1} $ and a relatively weak higher frequency components can be observed at $k_z=\pm 0.5\mu m ^{-1}$. \\

For widespread applicability of the proposed technique, we have also studied the response of the proposed optical system at varying illumination wavelengths. Fig.5 ($A_1, ~B_1, ~C_1$) shows the yz plane of the field distribution for illumination wavelengths, $364~nm , ~520~nm, ~630~nm$. The inset show a pitch of $180~nm , ~260~nm , ~320~nm$ with the width of each feature (nano-bump) of approximately $60~nm, ~120nm, ~140~nm$ respectively. Fig.$5A_2, ~B_2 , ~C_2$ correspond to the OTF at varying wavelengths. The spacing between the low and high freqiency components is quite evident. Moreover, the spacing is found to vary with illumination wavelength. This is clearly reflected in the adjoining intensity plots, Fig.$5A_3 ~B_3 , ~C_3$. This facilitates the scaling of distance between nano-bump and specific feature size. \\

In conclusion, we have proposed an efficient technique for fabricating periodic nano-structures. Unlike point-scanning techniques, we have employed interfering light-sheets configured in a $2\pi$ geometry. This has the advantage of high-throughput production when compared to point-scanning techniques. Phase matching is ensured for the counter-propagating light-sheets which upon superposition gives rise to nano-bump structures. The grating pattern thus generated has a feature (bump-width) of $60~nm$, and inter-bump spacing of $180~nm$. This technique will benefit mass production of nano-structures for applications in nano-electronics and nano-biology. \\\

\end{document}